\documentclass[12pt,preprint]{aastex}
\begin{document}
\newcommand{\calu}{{\cal U}}
\newcommand{\calq}{{\cal Q}}
\newcommand{\bx}{{\rm \bf x}}
\newcommand{\bk}{{\bar{\kappa}}}
\title{The Mass Function of Dark Matter Haloes in a Cosmological
Model with a Running Primordial Power Spectrum}
\author{Tong-Jie Zhang$^{1,2}$,Hong Liang$^{1}$,Bao-Quan Wang$^{3}
$,Wen-Zhong Liu$^{1}$}
\affil{$^1$Department of Astronomy, Beijing Normal University,
Beijing 100875, P.R.China; tjzhang@bnu.edu.cn \\
$^2$Canadian Institute for Theoretical Astrophysics(CITA),
University of Toronto,Toronto,ON M5S3H8,Canada;
tzhang@cita.utoronto.ca\\
$^3$Department of Physics, Dezhou University, De Zhou 253023,
P.R.China}

\begin{abstract}
We present the first study on the mass functions of Jenkins et al
(J01) and an estimate of their corresponding largest virialized
dark halos in the Universe for a variety of dark-energy
cosmological models with a running spectral index.
Compared with the PL-CDM model, the RSI-CDM model can raise the
mass abundance of dark halos for small mass halos at lower
redshifts, but it is not apparent on scales of massive mass halos.
Particularly, this discrepancy increases largely with the decrease
of redshift, and the RSI-CDM model can suppress the mass abundance
on any scale of halo masses at higher redshift. As for the largest
mass of virialized halos, the spatially flat $\Lambda$CDM models
give more massive mass of virialized objects than other models for
both of PL-CDM and RSI-CDM power spectral indexs, and
the RSI-CDM model can enhance the
mass of largest virialized halos for all of models considered in
this paper. So we probably distinguish the PL-CDM and RSI-CDM
models by the largest virialized halos in the future survey of
cluster of galaxies.

\end{abstract}


\section{Introduction}

The central problem in modern cosmology is the formation of large
scale structures in the universe. In the standard picture of
hierarchical structure formation, dark matter dominates the
universe, and a wide variety of observed structures, such as
galaxies, groups and clusters of galaxies, have formed by the
gravitational growth of Gaussian primordial density fluctuations.
Due to self-gravitational instability, the fluctuations of dark
matter have collapsed and virialized into objects which are
so-called `dark matter halos' or `dark halos'. The larger halos
are generally considered to have formed via the merger of smaller
ones collapsed first. The distribution of mass in the
gravitationally collapsed structures, such as galaxies and groups
(or clusters) of galaxies, which is usually called the mass or
multiplicity function, has been determined by observation.

As the observational data relevant to these issues improve, the
need for accurate theoretical predictions increases. By far the
most widely used analytic formulae for halo mass functions are
based on extensions of the theoretical framework first sketched by
\cite{1974ApJ...187..425P}. The Press-Schechter (PS) model theory
did not draw much attention until 1988, when the first relative
large N-Body simulation revealed a good agreement with it. The
mystery of the `fudge factor' of 2 in PS theory was solved by
approaching the 'cloud-in-cloud' problem with a rigorous
way\citep{1990MNRAS.243..133P,1991ApJ...379..440B}. The
reliability of the PS formula has been tested using N-Body
simulation by several authors, which turns out the PS formula
indeed provides an overall satisfactory description of mass
function for virialized objects. Unfortunately, none of these
derivations is sufficiently rigorous such that the resulting
formulae can be considered accurate beyond the regime where they
have been tested against N-body simulations. Although the
analytical framework of the PS model has been greatly refined and
extended in recent years, in particular to allow predictions for
the merger histories of dark matter halos
\citep{1991ApJ...379..440B}, it is well known that the PS mass
function, while qualitatively correct, disagrees in detail with
the results of N-body simulations. Specifically, the PS formula
overestimates the abundance of halos near the characteristic mass
and underestimates the abundance in the high mass tail. In order
to overcome this discrepancy, \cite{2001MNRAS.321..372J} proposed
an analytic mass function which gives a fit to their numerical
multiplicity function.

In particular, a power spectrum of primordial fluctuation,
$P_p(k)$, should be assumed in advance in the calculation of mass
function. Inflationary models predict a approximately
scale-invariant power spectra for primordial density (scalar
metric) fluctuation, $P_p(k)\propto k^n$ with index $n=1$
\citep{1982PhRvL..49.1110G,1983PhRvD..28..679B}. The combination
of the first-year Wilkinson Microwave Anisotropy Probe (WMAP) data
with other finer scale cosmic background (CMB) experiments (Cosmic
Background Imager [CBI], Arcminute Cosmology Bolometer Array
Receiver [ACBAR]) and two observations of large-scale structure
(the Anglo-Australian Telescope Two-Degree Field Galaxy Redshift
Survey [2dFGRS] and Lyman $\alpha$ forest) favour a $\Lambda$CDM
cosmological model with a running index of the primordial power
spectrum (RSI-$\Lambda$CDM), while the WMAP data alone still
suggest a best-fit standard power-law $\Lambda$CDM model with the
spectral index of $n\approx 1$ (PL-$\Lambda$CDM)
\citep{2003ApJS..148..175S,2003ApJS..148..213P}. However, there
still exist the intriguing discrepancies between theoretical
predictions and observations on both the largest and smallest
scales. While the emergence of a running spectral index may
improve problems on small scales, there remain a possible
discrepancy on the largest angular scales. It is particularly
noted that the running spectral index model suppress significantly
the power amplitude of fluctuations on small scales
\citep{2003ApJS..148..175S,2003ApJ...598...73Y}. This imply a
reduction of the amount of substructure within galactic halos
\citep{2002PhRvD..66d3003Z}. \cite{2003ApJ...598...73Y} studied
early structure formation in a RSI-$\Lambda$CDM universe using
high-resolution cosmological N-body/hydrodynamic simulations. They
showed that the reduced small-scale power in the RSI-$\Lambda$CDM
model causes a considerable delay in the formation epoch of
low-mass minihalos ($\sim10^6 M_{\sun}$) compared with the
PL-$\Lambda$CDM model, although early structure still forms
hierarchically in the RSI-$\Lambda$CDM model. Thus the running
index probably affect the abundance of dark halos formed in the
evolution of the universe.

Among the virialized structures, galaxy clusters are extremely
useful to cosmology because they may be in detail studied as
individual objects, and especially are the largest virialized
structure in the universe at present. The mass of a typical rich
clusters is approximately $10^{15}h^{-1}M_ {\odot}$, which is
quite similar to the average mass within a sphere of $8h^{-1}$Mpc
radius in the unperturbed universe. However, the theoretical
estimate of the mass of the largest collapsed object in the
RSI-$\Lambda$CDM cosmological framework has still not been
presented. Therefore, we will calculate the mass function of
collapsed objects by J01 mass functions respectively and present
the first calculation of the largest virialized object in the
Universe in a RSI-$\Lambda$CDM model to explore the effect of
running spectral index of primordial fluctuation on structure
formation.

The reminder of this paper is organized as follows. We describe
mass function of dark halos in Section 2. The largest virialized
dark halos in the universe are presented in Section 3. The
conclusion and discussion are given in Section 4.

\section{Mass Function of Dark Halos}

In the standard hierarchical theory of structure formation, the
comoving number density of virialized dark halos per unit mass $M$
at redshift $z$ can be expressed as: $n(M,z)=dN/dM=\rho_0
f(M,z)/M$ where $\rho_0$ is the mean mass density of the universe
today and, instead of PS formula in this letter, the mass function
$f(M,z)$ takes the form of an empirical fit from high-resolution
simulation \citep{2001MNRAS.321..372J}
\begin{equation}
f(M,z)=\frac{0.301}{M}\frac{d\ln\sigma^{-1}(M,z)}{d\ln
M}\exp(-|\ln\sigma^{-1}(M,z)+0.64|^{3.88}). \label{jenkins}
\end{equation}
Here $\sigma(M,z)=\sigma(M)D(z)$ and
$D(z)=e(\Omega(z))/e(\Omega_{\mathrm{m}})(1+z)$ is the linear
growth function of density perturbation
\citep{1992ARA&A..30..499C}, in which
$e(x)=2.5x/(1/70+209x/140-x^2/140+x^{4/7})$ and
$\Omega(z)=\Omega_{\mathrm{m}}(1+z)^3/ E^2(z)$. The present
variance of the fluctuations within a sphere containing a mass $M$
can be expressed as
$\sigma^2(M)=\frac{1}{2\pi^2}\int^{\infty}_0P(k)
W^2(kr_{\mathrm{M}})k^2dk$, where
$W(kr_{\mathrm{M}})=3[\sin(kr_{\mathrm{M}})/(kr_{\mathrm{M}})^3-
\cos(kr_{\mathrm{M}})/(kr_{\mathrm{M}})^2]$ is the Top-hat window
function in Fourier space and $
r_{\mathrm{M}}=(3M/4\pi\rho_0)^{1/3}$. The power spectrum of CDM
density fluctuations is $P(k)=P_p(k)T^2(k)$ where the matter
transfer function $T(k)$ is given by \cite{1999ApJ...511....5E},
and $P_p(k)$ is the primordial power spectrum of density
fluctuation. The scale-invariant primordial power spectrum in the
PL-$\Lambda$CDM model is given by $P_p(k)=Ak^{n_s}$ with index
$n_s$=1 and that in the RSI-$\Lambda$CDM model is assumed to be
$P_p(k)=P(k_0)(k/k_0)^{n_{s}(k)}$, where the index $n_{s}(k)$ is a
function of length scale
\begin{equation}
n_{s}(k)=n_{\rm s}(k_0) +\frac{1}{2}\frac{{\rm d}n_{s}(k)}{{\rm
d}\ln k}\ln \left(\frac{k}{k_0}\right).
\end{equation}
The pivot scale $k_0$=0.05 h Mpc$^{-1}$, $n_{\rm s}(k_0)$=0.93,
and $dn_{s}/d\ln k$=-0.03 are the best-fit values to the
combination data of the recent CMB experiments and two other
large-scale structure observations \citep{2003ApJS..148..175S}.
For both PL-$\Lambda$CDM and RSI-$\Lambda$CDM models, the
amplitude of primordial power spectrum, $A$ and $P(k_0)$, are
normalized to $\sigma_8=\sigma
(r_{\mathrm{M}}=8h^{-1}\mathrm{Mpc})$, which is the rms mass
fluctuations when present universe is smoothed using a window
function on a scale of $8h^{-1}\mathrm{Mpc}$. In this section, we
assume spatially flat $\Lambda$CDM models characterized by the
matter density parameter $\Omega_{\mathrm m}$, vacuum energy
density parameter $\Omega_{\Lambda}$. For both PL-$\Lambda$CDM and
RSI-$\Lambda$CDM models, we take cosmological parameters to be the
new result from the WMAP: Hubble constant $h=0.71$,
$\Omega_{\mathrm m}=0.27$, $\sigma_8=0.84$
\citep{2003ApJS..148....1B,2003ApJS..148..175S}.

\begin{figure}
\epsscale{0.8} \plotone{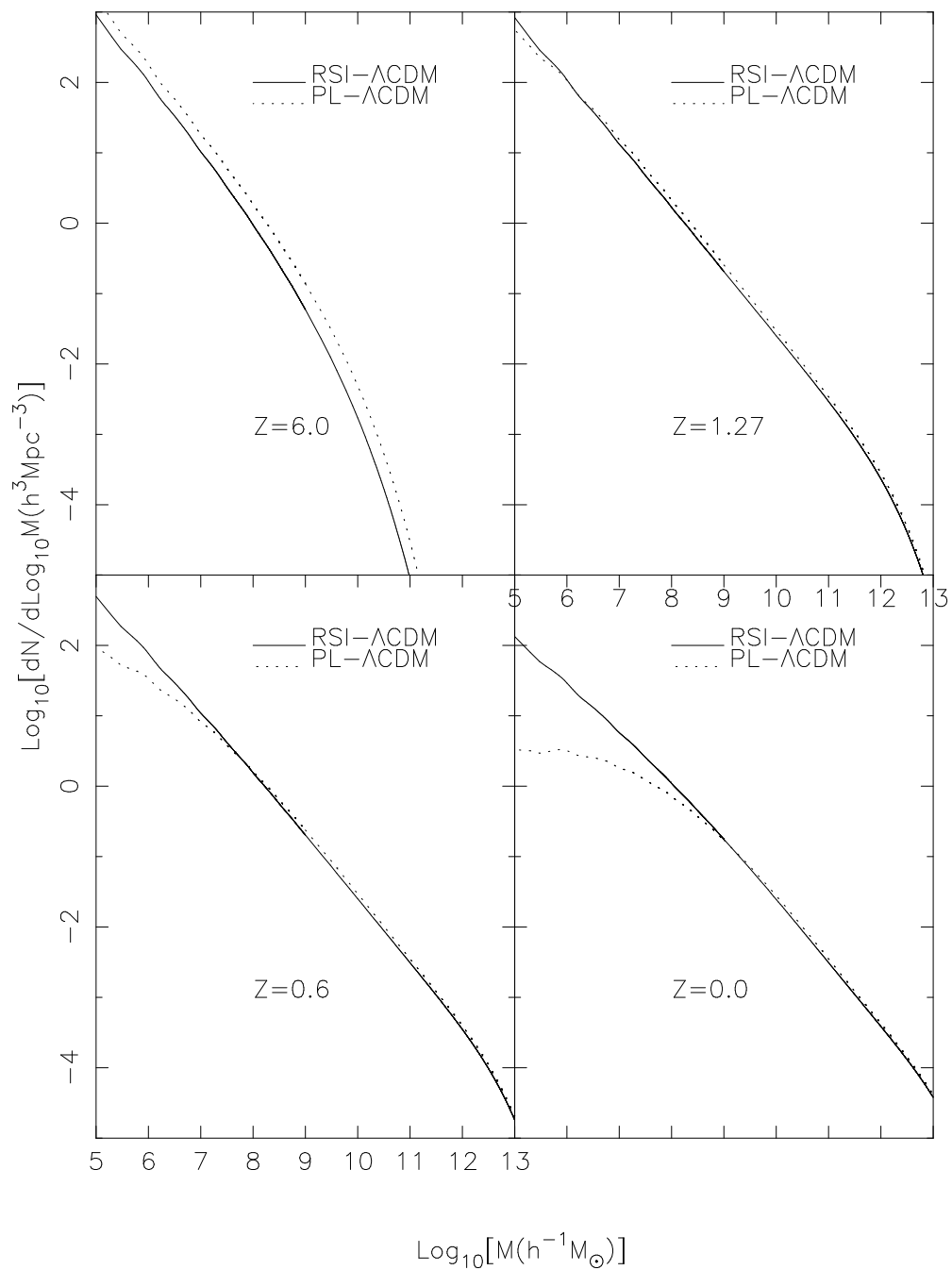} \caption{Mass function at redshift
$z$=0, 0.6, 1.27 and 6, respectively. The solid line is the mass
function for running spectral index $\Lambda$CDM model, while the
dashed one is that for power law $\Lambda$CDM model.}
\label{f1eps}
\end{figure}

The mass function of dark halos directly involve the calculation
of primordial power of density fluctuation. In order to explore
the difference between the two kinds of primordial power spectrum,
we first calculate the mass function of dark matter halos in a
wide range of redshift, which are plotted in Fig.(\ref{f1eps}). It
is noted that there is a slight difference between the PL-CDM
model and RSI-CDM model at lower redshifts. Compared with the
PL-CDM model, the RSI-CDM model can raise the mass abundance of
dark halos for small mass halos at lower redshifts, but it is not
apparent on scales of massive mass halos. Particularly, this
discrepancy increases largely with the decrease of redshift.
Similar to the result\citep{2003ApJ...598...73Y}, the
RSI-$\Lambda$CDM model can suppress the mass abundance on any
scale of halo masses at higher redshift. According to the
hierarchical formation theory of structure, there is fewer higher
mass halos at higher redshift and the higher mass halos are formed
by the merger of lower mass haloes at the relative late stage. As
pointed previously, the RSI model can suppress the power spectrum
at small scale, so this just leads to a considerable delay in the
formation of low mass haloes instead of high mass haloes.
Therefore, compared with the PL-CDM model, the mass function is
uniformly lower for the RSI model at higher redshift of $z$=6.

\section{The Largest Virialized Dark Halos In the Universe}

Based on the theoretical expression above, we can easily get the
total number $N$ of the virialized objects with the mass larger
than $M$
\begin{equation}
N=\int_{0}^{\infty}[\int_{M}^{\infty}\frac{dn}{dM}{\rm
d}M]\frac{dV}{dz}{\rm d}z, \label{eq:nv}
\end{equation}
where $dV$ is the comoving volume
element for the Friedmann-Robertson-Walker metric and
$\frac{dV}{dz}$ takes the form

\begin{eqnarray}
\frac{dV}{dz}= \left \{
\begin{array}{ll}
4\pi\frac{c^{3}}{H_{0}^{3}}\frac{D_{a}^{2}(1+z)^{2}\cos[\sqrt{\Omega_{k}}
\cdot f]}{\sqrt{1+\Omega_{k}D_{a}^{2}(1+z)^{2}}E(z)}
& for{\hspace{3mm}}\Omega_{k}<0 \\
4\pi\frac{c^{3}}{H_{0}^{3}}\frac{D_{a}^{2}(1+z)^{2}}{E(z)}&
for{\hspace
{3mm}}\Omega_{k}=0 \\
4\pi\frac{c^{3}}{H_{0}^{3}}\frac{D_{a}^{2}(1+z)^{2}\cosh[\sqrt{\Omega_{k}}
\cdot f]}{\sqrt{1+\Omega_{k}D_{a}^{2}(1+z)^{2}}E(z)}&
for{\hspace{3mm}}
\Omega_{k}>0,\\
\end{array}
\right.
\end{eqnarray}
where $D_{a}=d_{A}H_{0}/c$, $d_{A}$ is the angular diameter
distance and $f=\int_{0}^{z}{\rm d}z/E(z)$.

\begin{deluxetable}{rrrrrr}
\tablecolumns{6} \tablewidth{0pc} \tablecaption{Cosmological
Models Parameters} \tablehead{ \colhead{Model} &
\colhead{$\Omega_{m}$} & \colhead{$\Omega_{\Lambda}$} &
\colhead{$\Gamma$} & \colhead{$\sigma_8$}} \startdata
SCDM & 1 & 0 & 0.5& 0.6\\
LCDM & 0.3 & 0.7 & 0.21 & 1.0\\
OCDM & 0.3 & 0. & 0.25 & 1.0\\
\label{table:cm}
\enddata
\end{deluxetable}

It is obvious from the Eq.(\ref{eq:nv}) that the total number $N$
decrease with the increase of the mass $M$. Setting $N=1$, we can
finally obtain the largest mass $M_{MAX}$ of virialized object
\begin{equation}
1=\int_{0}^{\infty}[\int_{M_{MAX}}^{\infty}\frac{dn}{dM}{\rm
d}M]\frac{dV}{dz}{\rm d}z. \label{eq:nvmax}
\end{equation}
In this section, we consider three cold matter (CDM) models, i.e.
the standard CDM (SCDM), spatially flat $\Lambda$CDM models, and
an open CDM (OCDM)for both PL and RSI spectrum models. The
cosmological models parameters are given in Table \ref{table:cm}.
Then we calculate the largest virial mass $M_{MAX}$ in a variety
of cosmological models for both PL and RSI power spectrum model,
the results of which are demonstrated in Table \ref{table:nr}.
From Table \ref{table:nr} we can see that the different
cosmological models may yield the different result about virial
mass for the largest virialized halos. The spatially flat
$\Lambda$CDM models give more massive mass of virialized objects
than other models for both of PL and RSI power spectral models.
Therefore, it can distinguish different cosmological models by the
largest mass of virialized halos. Due to the accumulative effect
of the integration for volume(or redshift) over the whole space in
the universe, the prediction for virial mass is slightly greater
than the observed typical one. In addition, we also notice that
the RSI-CDM model can enhance the mass of largest virialized halos
for all of models considered here.

\begin{deluxetable}{rrrrrr}
\tablecolumns{6} \tablewidth{0pc} \tablecaption{Numerical Results
for the largest virial mass $M_{MAX}$($10^{15}h^{-1}M_{\odot}$)}
\tablehead{ \colhead{Model} & \colhead{PL model} & \colhead{RSI
model}} \startdata
SCDM & 3.0 & 3.3\\
LCDM & 6.2 & 6.9\\
OCDM & 4.5 & 5.05\\
\label{table:nr}
\enddata
\end{deluxetable}

\section{Conclusions and Discussion}

Motivated by the new result on the index of primordial power
spectrum from a combination of WMAP data with other finer scale
CMB experiments and other large-scale structure observations, we
present the first study on the mass functions of J01 and their
corresponding largest virialized dark halos in the Universe for a
variety of dark-energy cosmological models with a running spectral
index. It is well known that structures in the universe forms
hierarchically in standard CDM models. The most massive structure
form rather late in the universe. It is also noted that there is a
slight difference between the mass abundance of PL-CDM and RSI-CDM
model at lower redshifts. Compared with the PL-CDM model, the
RSI-CDM model can raise the mass abundance of dark halos for small
mass halos at lower redshifts, but it is not apparent on scales of
massive mass halos. Particularly, this discrepancy increases
largely with the decrease of redshift, and the RSI-$\Lambda$CDM
model can suppress the mass abundance on any scale of halo masses
at higher redshift.

As for the largest mass of virialized halos, the spatially flat
$\Lambda$CDM models give more massive mass of virialized objects
than other models for both of PL-CDM and RSI-CDM power spectral
models. Therefore, it can distinguish different cosmological
models by the largest mass of virialized halos for both of PL-CDM
and RSI-CDM models.
In addition, we also notice that the RSI-CDM model can enhance the
mass of largest virialized halos for all of models considered
here. So we probably distinguish the PL-CDM and RSI-CDM models by
the largest virialized halos in the future survey of cluster of
galaxies.
Therefore, the obtained largest virialized object can be referred
to as the complement to the observations of CMB, SN Ia and large
scale structure in the future cosmological observation.


\cite{2003ApJ...598...73Y} found that although the hierarchical
formation mechanism do not work well in RSI-$\Lambda$CDM model
compared with that in PL-$\Lambda$CDM model and it also is not
clear that the PS theory can be used in RSI-$\Lambda$CDM model,
the mass function measured by high-resolution cosmological
N-body/hydrodynamic simulations overall match the PS mass function
for both RSI-$\Lambda$CDM and PL-$\Lambda$CDM model. In addition,
because the running spectral index model predicts a significant
lower power of density fluctuation on small scales than the
standard PL-$\Lambda$CDM
model\citep{2003ApJS..148..175S,2003ApJ...598...73Y}, it should
also attract considerable attention in studies on strong lensing
\citep{2004astro.ph.10431Z,2003ApJ...587L..55C,2003A&A...397..415C,
2004ChJAA...4..118C,2004A&A...418..387C, 2004ApJ...602L...5Z} and
weak lensing by large-scale structure\citep{2004PhRvD..69h3514I},
especially on
skewness\citep{2003ApJ...592..664P,2003ApJ...598..818Z,2005astro.ph..3064Z},
which characterizes the non-Gaussian property of $\kappa$ field in
the nonlinear regime.

\acknowledgments
We are very grateful to the anonymous referee for
valuable comments that greatly improved the paper. We also thank Ue-Li Pen 
for useful discussions. T.J.Zhang would
like to thank CITA for its hospitality during his visit. This work
was supported by the National Science Foundation of China (Grants
No.10473002 and 10273003), the Scientific Research Foundation for
the Returned Overseas Chinese Scholars, State Education Ministry.

\bibliography{ztjcos-lensbib}
\bibliographystyle{apj}

\appendix

\end{document}